\documentclass[aps,manuscript]{revtex4}
\begin{document}

\title{Comment on ``The complete Schwarzschild interior and exterior solution 
in the harmonic coordinate system''\\ {[}J. Math. Phys. 39, 6086 (1998){]}}

\author{L\'{a}szl\'{o} \'{A}. Gergely}

\address{Laboratoire de Physique Th\'{e}orique, Universit\'{e}
Louis Pasteur,\\ 3-5 rue de l'Universit\'{e} 67084 Strasbourg Cedex, France\dag
\\
and\\
KFKI Research Institute for Particle and Nuclear Physics,\\
Budapest 114, P.O.B 49, H-1525 Hungary}

\maketitle
  
\newpage
 
In a recent paper Liu \cite{Liu} considered the complete Schwarzschild  
interior and exterior solution in harmonic coordinates. There he argued 
about the necessity to keep the integration constant $C_1$ in $R_{ex}$, 
in contrast with previous treatments (Refs. 1-5 and Ref. 8 of Ref. 1).
The purpose of this comment is to show that the above conclusion cannot be 
traced from the matching conditions between the vacuum exterior and the 
uniform density interior perfect fluid, as claimed in \cite{Liu}.  
The reason for this is that the last condition in Eqs. (7) 
of Ref. 1, namely ${R'}_{in}(a)={R'}_{ex}(a)$ is {\em not} required by 
the junction conditions at $r=a$, as will be shown in what follows.

The junction of two space-times along a timelike hypersurface $\Sigma$ 
can be done applying the Darmois-Israel matching procedure 
\cite{Darmois,Israel}, which requires the continuity across the junction 
of both the first and second fundamental forms (induced metric and 
extrinsic curvature) {\em of the junction hypersurface}. 

In the standard coordinates, the metric (1) of Ref. 1 induces 
the 3-metric given by the line element
\begin{equation}
ds^2_\Sigma=E(r)dt^2-r^2(d\theta^2+\sin^2\theta d\phi^2)  
\label{firstform}
\end{equation}
on the junction hypersurface $r=a$, which has the normal 
$n=1/\sqrt{G}\ \partial /\partial r$ and the nonvanishing 
extrinsic curvature components
\begin{equation}
K_{00}=-\frac{E'}{2\sqrt{G}}\ ,\qquad
K_{22}=\frac{r}{\sqrt{G}}\ ,\qquad
K_{33}=\frac{r}{\sqrt{G}}\sin^2\theta\ .
\label{secondform}
\end{equation}
The continuity of both of (\ref{firstform}) and (\ref{secondform})
implies that the metric function $G$ is continuous and $E$ is ${\cal C}^1$ 
across the junction.
It is not to be expected that starting from the same metrics written in other 
coordinate systems the conditions on $E$ and $G$ will be weakened.
However, constraints on the new function $R$ introduced by the coordinate 
transformation (2)-(5) of Ref. 1 will also emerge. 

In the harmonic coordinate system the normal vector to $\Sigma$ has the 
components $n^\mu=(\ln R)'/\sqrt{G}\ (0,X_1,X_2,X_3)$.
The metric (6) of Ref. 1 (with a missing square on the last bracket 
corrected) induces the first fundamental form
\begin{equation}
ds^2_\Sigma=E(r)dt^2-\frac{r^2}{R^2(r)}d{\bf X}^2\ . 
\label{firstformharm}
\end{equation} 
This is still expressed in terms of the four space-time coordinates.
When written in terms of the coordinates $t,\theta$ and $\phi$, intrinsic 
to $\Sigma$, the continuity of the induced metric again implies the continuity 
of the metric function $E$ alone. The extrinsic curvature tensor in the new 
coordinate system is found either by direct computation or by transforming 
its components (\ref{secondform}) from standard to harmonic coordinates.
The nonvanishing components are $K_{00}$ given in (\ref{secondform}) and 
\begin{equation}
K_{ii}=\frac{(X_j^2+X_k^2)r}{R^4\sqrt{G}}\ ,\qquad
K_{ij}=-\frac{X_iX_jr}{R^4\sqrt{G}}\ ,
\label{secondformharm}
\end{equation}
where $i\neq j\neq k$ take the values 1,2,3.
The junction condition on the extrinsic curvature at arbitrary radius
implies that $G, R$ and $E'$ should be continuous. Altogether we find 
that $G$ and $R$ are ${\cal C}^0$ and $E$ is ${\cal C}^1$. Thus the last 
relation in (7) of Ref. 1 does not hold. 

The condition 
\begin{equation}
{E'}_{in}={E'}_{ex}
\ , \label{extracond}
\end{equation} 
though not listed among the continuity 
conditions (7) of Ref. 1, was fulfilled when imposing that the pressure
vanishes on the junction. Indeed, Eq. (\ref{extracond}) is a substitute 
for the requirement that the radial pressures on the two sides of $\Sigma$ 
are equal, which was demonstrated for generic spherically symmetric static 
space-times in an other context \cite{GergelyND}. 

Our criticism does not affect the main result of Ref. 1, which is 
the solution $R_{in}$ of the second order differential equation (22) 
of Ref. 1. The arguments about the integration constants however should be
reviewed. Requiring only the continuity of $R$ and nothing more, one of
the constants $C_1$ and $C_2$ can be freely specified, in particular
$C_1=0$ can be chosen, in accordance with Refs. 1-5 and Ref. 8 of Ref. 1.
Of course, the continuity of $R'$ across the junction of the interior
and exterior Schwarzschild solutions can be imposed as an additional
requirement for other purposes (e.g. for having a smooth function $R(r)$ 
as in Ref. 1), but it is not a consequence of the junction conditions.

{\dag \ visiting position, supported by the Hungarian State 
E\"{o}tv\"{o}s Fellowship}
 
\end{document}